\documentclass[12pt]{iopart}

\usepackage{iopams}
\usepackage{graphicx}
\usepackage{color}

\eqnobysec

\newcommand{\Real}{\mathrm{Re}\,}
\newcommand{\Imag}{\mathrm{Im}\,}
\newcommand{\order}{\mathcal{O}}

\newcommand{\R}{\mathbb{R}}
\newcommand{\Z}{\mathbb{Z}}

\begin{document}

\title{Travelling kinks in discrete $\phi^4$ models}

\author{O F Oxtoby\dag, D E Pelinovsky\ddag\ and I V Barashenkov\S
\footnote[5]{On sabbatical leave from the University of Cape Town,
South Africa}}
\address{\dag\ Department of Mathematics and Applied Mathematics, University of Cape Town,
Rondebosch 7701, South Africa}
\address{\ddag\ Department of Mathematics, McMaster
University, Hamilton, Ontario, Canada, L8S 4K1}
\address{\S\ Physikalisches Institut, Universit\"at Bayreuth, D-95440 Bayreuth, Germany.}
\eads{\mailto{olivero@maths.uct.ac.za}, \mailto{dmpeli@math.mcmaster.ca} and
\mailto{igor@cenerentola.mth.uct.ac.za}}

\begin{abstract}
In recent years, three exceptional discretizations of the $\phi^4$ theory
have been discovered \cite{Speight1, BT97, Panos} which support translationally
invariant kinks, i.e. families of stationary kinks centred at arbitrary
points between the lattice sites. It has been suggested that the translationally
invariant stationary kinks may persist as {\it sliding\/}
kinks, i.e. discrete kinks travelling at nonzero velocities without experiencing
any radiation damping. The purpose of this study is to check
whether this is indeed the case. By computing the Stokes constants
in beyond-all-order asymptotic expansions, we prove that the three
exceptional discretizations do not support sliding kinks for most
values of the velocity --- just like the standard, one-site, discretization.
There are, however, isolated values of velocity for which radiationless
kink propagation becomes possible. There is one such value for the
discretization of \cite{Speight1} and three {\it sliding velocities\/} for the
model of \cite{Panos}.
\end{abstract}


\section{Introduction}

Spatially discretized partial differential equations
(or, equivalently, chains of coupled ordinary differential equations) have
attracted considerable attention recently. One of the issues that has
been vigorously debated and that will concern us in this
paper, is whether discrete systems can support solitary waves
travelling without losing energy to resonant 
radiation and decelerating as a result. 
We address this issue for one of the prototype
models of nonlinear physics, the $\phi^4$-theory:
\begin{equation}
\label{continuous-phi4} u_{tt} = u_{xx} + \frac{1}{2} u ( 1 - u^2).
\end{equation}
The $\phi^4$-equation (\ref{continuous-phi4}) is
Lorentz-invariant, and so the existence of the travelling kink
\begin{equation}
\label{travelling} u(x,t) = \tanh \frac{x - ct - s}{2 \sqrt{1 -
c^2}},
\end{equation}
where $|c| < 1$ and $s \in \mathbb{R}$, is an immediate
consequence of the existence of the stationary kink for $c=0$. On
the other hand, if we discretize equation (\ref{continuous-phi4}) in
$x$,
\begin{equation}
\label{phi-4} \ddot{u}_n = \frac{u_{n+1} - 2 u_n + u_{n-1}}{h^2} +
f(u_{n-1},u_n,u_{n+1}),
\end{equation}
the translation and Lorentz invariances are lost, and the existence
of the travelling kink (and even of an arbitrarily centred stationary one)
 becomes a
nontrivial matter. In equation (\ref{phi-4}), $u_n \in \R$, $n \in \Z$,
$t \in \R$, $h$ is the lattice spacing, and the nonlinearity
$f(u_{n-1},u_n,u_{n+1})$ satisfies the continuity condition
\begin{equation}
\label{nonlinearity-continuous} f(u,u,u) = \frac{1}{2} u (1 - u^2).
\end{equation}
We restrict ourselves to symmetric discretizations, i.e.                                       
\begin{equation}
\label{nonlinearity-symmetry} f(u_{n-1},u_n,u_{n+1}) =
f(u_{n+1},u_n,u_{n-1}).
\end{equation}
Equation (\ref{continuous-phi4}) results from (\ref{phi-4}) in the
continuum limit, where $u_n(t)=u(x_n,t)$, $x_n=nh$ and $h \to 0$.
In this limit, the truncation error of the Taylor series
is $\order(h^2)$. We shall be concerned with monotonic kink solutions of                          
(\ref{phi-4}):  $u_{n+1}(t) \geq u_n(t)$  for all $n \in \mathbb{Z}$. As $h \to 0$, such monotonic 
discrete kinks approach the continuous
kink (\ref{travelling}).

The most common, one-site discretization of the nonlinearity function is
given by
\begin{equation}
\label{nonlinearity-phi4}
f(u_{n-1},u_n,u_{n+1}) = \frac{1}{2} u_n(1 - u_n^2).
\end{equation}
It is a well  established fact \cite{Lazutkin}, however, that the
discrete Klein-Gordon equation
(\ref{phi-4})+(\ref{nonlinearity-phi4}) admits only a countable
set of stationary monotonic kinks with the boundary conditions
\begin{equation}
\label{BCs} 
\lim_{n \to -\infty} u_n(t) = -1, \qquad \lim_{n \to
+\infty} u_n(t) = +1.
\end{equation}
Physically, this fact is related to the
presence of the Peierls-Nabarro barrier, an effective potential
periodic with the spacing of the lattice. Half of the stationary
kinks are centred at the minima (the on-site kinks) and the other half
(the off-site kinks) at the maxima of the Peierls-Nabarro potential. There are no
continuous families of stationary discrete kinks of the form
$u_n=u(n-s)$, with $s$ a free parameter,  which would interpolate between the two solutions. 
In an abuse of terminology, we will be calling such families ``translationally
invariant kinks'' --- although, in the first place,  translation invariance 
is a property of an equation rather than a solution, and in the second,
all lattice equations are of course \emph{not} translationally invariant.
As for propagating waves, of special importance are kinks moving at constant speed and 
without the emission of radiation. We will be referring to such kinks, i.e. solutions of the form
$u_n=u(n-ct-s)$ where $u(\xi)$ is a monotonically
growing function satisfying the boundary conditions (\ref{BCs}),  as
{\it sliding kinks\/}, to emphasise the fact that they do not 
 experience any radiative friction.
Being an obstacle to the ``translationally
invariant" kinks, the Peierls-Nabarro barrier is also detrimental to 
the existence of sliding
kinks --- at least for small $c$ (see reviews in \cite{S03}
and \cite{IJ04}).

In an attempt to find a discrete model 
with ``translationally invariant" and sliding kinks, 
Speight and Ward \cite{Speight1,Speight2} considered a hamiltonian discretization
of the form
\begin{eqnarray}
\nonumber f(u_{n-1},u_n,u_{n+1}) & = & \frac{1}{12} (2 u_n +
u_{n+1}) \left( 1 - \frac{u_n^2 + u_n u_{n+1} + u_{n+1}^2}{3}
\right) \\
\label{nonlinearity-Speight} & + & \frac{1}{12} (2 u_n + u_{n-1})
\left( 1 - \frac{u_n^2 + u_n u_{n-1} + u_{n-1}^2}{3} \right).
\end{eqnarray}
In the static limit, the corresponding energy admits a topological
lower bound which is
saturated by a first- (rather than second-) order difference
equation.  This equation is readily shown to have a
one-parameter continuous family of stationary kink solutions $u_n=u(n-s)$
for $0 \le h \le 2$ (see Proposition 1 in
\cite{Speight2}). The parameter $s$ of the family defines the
position of the kink relative to the lattice. Since all members of the family have the
same (lowest attainable) energy, the stationary kink experiences no
Peierls-Nabarro barrier. As for travelling kinks, Speight and Ward's numerical
simulations revealed that although moving kinks in
this model do lose energy to Cherenkov radiation and decelerate as
a result, this happens at a slower rate than a similar process in
equation (\ref{nonlinearity-phi4}) (see figures 4 and 5 in
\cite{Speight2}).

Another line of attack was chosen by Bender and Tovbis \cite{BT97} who proposed
a different discretization supporting a continuous family
of arbitrarily centred stationary kinks:
\begin{equation}
\label{nonlinearity-Tovbis} f(u_{n-1},u_n,u_{n+1}) = \frac{1}{4}
\left( u_{n+1} + u_{n-1} \right) \left( 1 - u_n^2 \right).
\end{equation}
In this case, the family arises due to the suppression of the stationary
kink's resonant radiation. In fact, the family of stationary kinks
can be found explicitly as
\begin{equation}
\label{AL-kink} u_n(t) = \tanh \left[ a (n-s) \right],
\end{equation}
where $a = {\rm arcsinh}\left( h/2 \right)$ for all $h \in
\mathbb{R}$. (The solution (\ref{AL-kink}) coincides with the stationary
dark soliton of the repulsive Ablowitz-Ladik equation \cite{AH93}.)

Finally, the nonlinearity
\begin{equation}
\label{nonlinearity-Panos} f(u_{n-1},u_n,u_{n+1}) = \frac{1}{8}
\left( u_{n+1} + u_{n-1} \right) \left( 2 - u_{n+1}^2 - u_{n-1}^2
\right).
\end{equation}
was introduced by Kevrekidis \cite{Panos}, who demonstrated the
existence of a two-point invariant and hence a first-order
difference equation associated with the stationary equation.
Consequently, 
the discretization (\ref{nonlinearity-Panos}) also
supports a continuous family of stationary kinks for all $h \in [0, h_0]$ with some
$h_0 > 0$. (For general discussion, see \cite{Speight3}, \cite{BOP05} and
\cite{DKY05a}.)
A relevant property of the model (\ref{nonlinearity-Panos}),
which is related to the existence of a two-point invariant \cite{Panos}
and indicates some additional underlying symmetry,
is the conservation of momentum.  (See also \cite{DKY05b}.)

Since the reasons for the nonexistence of ``translationally invariant''
kinks and of sliding kinks are apparently related
(the breaking of symmetries of the underlying continuum theory or,
speaking physically, the presence of the Peierls-Nabarro barrier),
the availability of ``translation-invariant"  stationary kinks in the models
(\ref{nonlinearity-Speight}), (\ref{nonlinearity-Tovbis}) and
(\ref{nonlinearity-Panos}) suggests that they might have sliding kinks
as well. 
It is the purpose of the
present study to find out whether this is indeed the case. We shall analyse the persistence of continuous families
of stationary kinks $u_n=u(n-s)$ for nonzero velocities; in other
words, examine the existence of solutions
of the form $u(n-ct-s)$ where $u(z)$ is a monotonically
growing function satisfying (\ref{BCs}),  and $c
\neq 0$. We develop an accurate numerical test in the limit $h \to 0$ 
which shows whether or not standing and travelling kinks of the
discrete $\phi^4$ model (\ref{phi-4}) bifurcate from the exact kink
solutions (\ref{travelling}) of its continuous counterpart
(\ref{continuous-phi4}). The analysis of this bifurcation poses a 
singular problem in perturbation theory 
which can be analysed using two (inner and outer) matched asymptotic
scales on the complex plane \cite{Tovbis1,Tovbis2}. 
In particular, the nonvanishing of the Stokes constant in the
inner asymptotic equation serves as a sufficient condition
for the non-existence of continuous solutions of the
difference equations \cite{Tovbis1}.

Our test will be based on computing  the Stokes constant
for the differential-difference equation underlying the lattice system.
We will examine  all four 
discretizations of the $\phi^4$ theory mentioned above,
i.e. equations (\ref{nonlinearity-phi4}), (\ref{nonlinearity-Speight}),
(\ref{nonlinearity-Tovbis}) and (\ref{nonlinearity-Panos}). 
Since translationally invariant stationary kinks $u_n = u(n-s)$ do exist
for the three exceptional nonlinearities (\ref{nonlinearity-Speight}),
(\ref{nonlinearity-Tovbis}) and (\ref{nonlinearity-Panos}),
the Stokes constant
is \emph{a priori} vanishing for $c=0$ in these three cases. However, 
we will show that in all three cases the Stokes constant acquires a nonzero 
value as soon as $c$ deviates from zero. It remains nonzero for all $c$ 
except a few isolated values which define the particular velocities of the sliding
kinks in the corresponding model. There is one such isolated zero of the Stokes
constant for the nonlinearity (\ref{nonlinearity-Speight}) and three
{\it sliding velocites\/} for the discretization (\ref{nonlinearity-Panos}). 
Consequently, the main conclusion of this work is that the sliding 
kinks,
 i.e.   kinks travelling at a constant speed without the emission of radiation,
can occur only at particular values of the velocity.
The sliding velocities are, of course, functions of the 
discretization spacing $h$, so that sliding kinks arise 
along continuous curves on the $(c,h)$-plane.

We conclude this introduction with a remark on a convention
adopted in the remainder of this paper --- namely, that the 
linear part of the function
$f(u_{n-1},u_n,u_{n+1})$ in (\ref{phi-4}) can always be fixed to
$\frac12 u_n$ without loss of generality.  Indeed, the most general
function satisfying (\ref{nonlinearity-continuous}) and
(\ref{nonlinearity-symmetry}) is $f = \left( \frac{1}{2} - 2
a\right) u_n + a \left(u_{n+1}+u_{n-1}\right) + \textrm{cubic terms}$,
where $a$ is arbitrary. Since $h^2$ in (\ref{phi-4}) is also a free
parameter, we can always make a replacement $h\to\tilde{h}$ such that
$1/h^2 + a = 1/\tilde{h}^2$.  This gives
\begin{equation}
f(u_{n-1}, u_n, u_{n+1}) = \frac12 u_n - Q(u_{n-1}, u_n, u_{n+1}), \label{canonical-form}
\end{equation}
where $Q$ is a homogeneous polynomial of degree 3 which is
independent of the parameter $h$.

The plan of this paper is as follows.
In the next section (section \ref{section2}) we review the construction
of the outer and inner asymptotic solutions in the limit $h \to
0$. Section \ref{section3} contains details of the numerical computation
of the Stokes constants while the last section (section \ref{section4})
summarises the results of our work.

\section{Inner and outer asymptotic expansions in the limit $h \to
0$}

\label{section2}

We are looking for a sliding-kink solution of the discrete $\phi^4$
models (\ref{phi-4}) in the form
\begin{equation}
\label{trav-wave} u_n(t) = \phi(z), \qquad z = h(n-s) - ct,
\end{equation}
where $\phi(z)$ is assumed to be a twice differentiable function of
$z \in \mathbb{R}$, that satisfies the differential advance-delay
equation
\begin{equation}
\label{advance-delay} \fl c^2 \phi''(z) = \frac{\phi(z + h) - 2
\phi(z) + \phi(z-h)}{h^2} + \frac{1}{2} \phi(z) -
Q\left(\phi(z-h),\phi(z),\phi(z+h)\right),
\end{equation}
with the boundary conditions $\phi(z) \to \pm 1$ as $z \to
\pm \infty$. The velocity $c$ is assumed to be smaller than 1
in modulus. If a solution to this boundary-value problem (i.e.
a heteroclinic orbit) exists, then the parameter $s$
is arbitrary due to the translation invariance of the advance-delay
equation (\ref{advance-delay}). The scaling parameter $h$ (which
stands for the lattice step-size) can be used to reduce equation (\ref{advance-delay})
to a singularly perturbed differential equation as $h \to 0$
\cite{Tovbis1}. Formal asymptotic solutions of the problem
(\ref{advance-delay}) can be constructed at the inner and outer
asymptotic scales. The formal series represent convergent asymptotic
solutions of the singular perturbation problem only if the Stokes
constants are all zero \cite{Tovbis2}.

Asymptotic analysis beyond all orders of perturbation theory was
pioneered by Kruskal and Segur \cite{SegurKruskal2} and has been
utilised by many authors. It was extended by Pomeau
et.~al.~\cite{PomeauRamani} to allow the computation of radiation
coefficients from Borel summation of series rather than from the
numerical solution of differential equations. Essentially the same
method has been applied to different problems by Grimshaw and Joshi
\cite{GrimshawJoshi, GrimshawNLS} and Tovbis and collaborators
\cite{Tovbis1, Tovbis2, Tovbis3,TovPel}. In this paper, we shall
work with formal inner and outer asymptotic series for the problem
(\ref{advance-delay}) without attempting rigorous analysis of their 
asymptoticity.

\subsection{Outer asymptotic series}

Assuming that the solution $\phi(z)$ is a real analytic function
of $z$, we consider the Taylor series for the second difference
in a strip ${\cal D}_{\delta} = \left\{ z \in \mathbb{C} : \;\;
|\Imag z| < \delta \right\}$, where $\delta > 0$:
\begin{equation}
\label{Taylor1} \phi(z+h) - 2 \phi(z) + \phi(z-h) = h^2 \phi''(z)
+ \sum_{n=2}^{\infty} h^{2n} \frac{2}{(2n)!} \phi^{(2n)}(z).
\end{equation}
Since the cubic polynomial $Q(u_{n-1},u_n,u_{n+1})$ satisfies the
continuity and symmetry relations (\ref{nonlinearity-continuous})
and (\ref{nonlinearity-symmetry}), the nonlinearity of
(\ref{advance-delay}) can also be expanded in a Taylor series in the
same strip:
\begin{equation}
\label{Taylor2} \fl Q\left(\phi(z-h),\phi(z),\phi(z+h)\right) =
\frac{1}{2} \phi^3(z) + \sum_{n=1}^{\infty} h^{2n}
Q_{2n}\left(\phi,(\phi')^2,...,\phi^{(2n)}\right),
\end{equation}
where the coefficients $Q_{2n}$ depend on even derivatives
and even powers of odd derivatives of $\phi(z)$, and also
$Q_{2n}(\phi,0,...,0) = 0$. The differential advance-delay equation
(\ref{advance-delay}) can thus be written as
\begin{eqnarray}
\label{singular-perturbation} \fl (1-c^2) \phi'' + \frac{1}{2} \phi ( 1
- \phi^2) + \sum_{n=1}^{\infty} h^{2n} \left( \frac{2}{(2n+2)!}
\phi^{(2n+2)} - Q_{2n} \left(\phi,(\phi')^2,\dots,\phi^{(2n)}\right)
\right) = 0.\nonumber\\
\end{eqnarray}
For $h = 0$, equation (\ref{singular-perturbation}) becomes the 
travelling wave reduction of the continuous model
(\ref{continuous-phi4}), with the explicit solution
\begin{equation}
\label{leading-order} \phi_0(z) = \tanh \xi; \quad \xi =
\frac{z}{2 \sqrt{1 - c^2}}, \quad |c| < 1.
\end{equation}

We will search for solutions of equation (\ref{singular-perturbation}) of the form
\begin{equation}
\label{perturbation-series} \hat{\phi}(z) = \phi_0(z) +
\sum_{n=1}^{\infty} h^{2n} \phi_{2n}(z).
\end{equation}
Substituting the expansion (\ref{perturbation-series}) into
(\ref{singular-perturbation}) we get, at order $h^{2n}$,
\[
\mathcal{L} \phi_{2n} = H_{2n},
\]
where the linearised operator ${\cal L}$ is given by
\[
\mathcal{L} = - \frac{d^2}{d \xi^2} + 4 - 6\,\mathrm{sech}^2\,
\xi,
\]
and $H_{2n}$ are polynomials in $\phi_0, \phi_2, \ldots,
\phi_{2n-2}$ and their derivatives. The kernel of
$\mathcal{L}$ is one-dimensional, and spanned by an even
eigenfunction $y_0 = \mathrm{sech}^2\, \xi$.  The rest of the
spectrum of $\mathcal{L}$ is positive. It is not difficult
to prove by induction that if $\phi_{2k}(z)$ are all odd in $z$ for
$k = 0,1,...,n-1$, the nonhomogeneous term $H_{2n}$ is also odd in
$z$ and hence, by the Fredholm alternative, there exists a unique
odd bounded solution $\phi_{2n}(z)$ for $z \in \mathbb{R}$.
Moreover, since $H_{2n}$ decays to zero exponentially fast as $|z|
\rightarrow \infty$, the function $\phi_{2n}(z)$ is also
exponentially decaying for any $n \ge 1$. The perturbation
$\phi_2(z)$ in particular satisfies the nonhomogeneous equation
\begin{equation}
\label{first-order} \mathcal{L} \phi_2 = - \frac{1}{3}
\left( \phi_0^{({\rm iv})} + \alpha \phi_0'' + \beta \phi_0^2
\phi_0'' + \gamma \phi_0 (\phi_0')^2 \right),
\end{equation}
where the numerical coefficients depend on whether the nonlinearity
function $f$ is given by (\ref{nonlinearity-phi4}), (\ref{nonlinearity-Speight}),
(\ref{nonlinearity-Tovbis}) or (\ref{nonlinearity-Panos}):
\begin{eqnarray*}
\textrm{One-site (\ref{nonlinearity-phi4})} : & \phantom{t} & \alpha = \beta = \gamma
= 0; \\
\textrm{Speight-Ward (\ref{nonlinearity-Speight})} : & \phantom{t} & \alpha = 1, \; \beta
= \gamma = -4; \\
\textrm{Bender-Tovbis (\ref{nonlinearity-Tovbis})} : & \phantom{t} & \alpha = 3, \; \beta =
-3, \; \gamma = 0; \\
\textrm{Kevrekidis (\ref{nonlinearity-Panos})} : & \phantom{t} & \alpha = 3, \; \beta =
-9, \; \gamma = -6.
\end{eqnarray*}
The odd bounded solution $\phi_2(z)$ of the nonhomogeneous
equation (\ref{first-order}) is:
\begin{equation}
\label{first-order-explicit} \phi_2(z) = {\cal A} \tanh \xi \,
\mathrm{sech}^2 \, \xi + {\cal B} \xi \, \mathrm{sech}^2\,\xi,
\end{equation}
where
\begin{eqnarray*}
{\cal A} = \frac{(1-c^2)(\gamma + 2 \beta) + 6}{72 (1-c^2)^2}, \qquad  
{\cal B} = -\frac{(1-c^2)(\alpha + \beta) + 1}{24 (1-c^2)^2}.
\end{eqnarray*}
The hat in the series (\ref{perturbation-series}) indicates that the
series is formal, i.e.~it may or may not converge
\cite{Tovbis1,Tovbis2}, depending on the choice of $c$ and $Q$ in
equation (\ref{advance-delay}). We shall be referring to
(\ref{perturbation-series}) as the outer asymptotic expansion.

\subsection{Inner asymptotic series}

The leading-order term (\ref{leading-order}) of the outer expansion
(\ref{singular-perturbation}) has poles at
$\xi = \frac{\pi \rmi}{2} (1 + 2n)$, where $n \in \mathbb{Z}$.
We apply the scaling transformation
\begin{equation}
\label{scaling} z = h \zeta + \rmi \pi \sqrt{1-c^2}, \qquad \phi(z) =
\frac{1}{h} \psi(\zeta)
\end{equation}
to equation (\ref{advance-delay}) in order to study the convergence of the formal
asymptotic solution (\ref{perturbation-series})
near the pole
$\xi = \frac{\pi \rmi}{2}$ (see \cite{Tovbis1,Tovbis2}).
This yields the following differential
advance-delay equation for $\psi(\zeta)$:
\begin{eqnarray}
\label{inner} \fl c^2 \psi''(\zeta) = \psi(\zeta +1) - 2\psi(\zeta) +
\psi(\zeta -1) -
Q\left(\psi(\zeta-1),\psi(\zeta),\psi(\zeta+1)\right) +
\frac{h^2}{2} \psi(\zeta).
\end{eqnarray}
The following are the cubic functions $Q$ for each of the four
discretizations that we deal with in this paper:
\begin{eqnarray*}
\fl \quad \textrm{One-site (\ref{nonlinearity-phi4})}: & Q &= \frac{1}{2}
\psi^3(\zeta); \\
\fl \quad \textrm{Speight-Ward (\ref{nonlinearity-Speight})}: & Q
&= \frac{1}{36} \left[ \psi^3(\zeta+1) + 3
\psi^2(\zeta+1)\psi(\zeta) + 3 \psi(\zeta+1)\psi^2(\zeta) \right. \\
& \phantom{Q} &\phantom{=} \left. + 4\psi^3(\zeta) + 3 \psi(\zeta-1)
\psi^2(\zeta) + 3 \psi^2(\zeta-1)\psi(\zeta) + \psi^3(\zeta-1) \right]; \\
\fl \quad \textrm{Bender-Tovbis (\ref{nonlinearity-Tovbis})}: & Q &=
\frac{1}{4} \psi^2(\zeta) \left[ \psi(\zeta+1) + \psi(\zeta-1) \right]; \\
\fl \quad \textrm{Kevrekidis (\ref{nonlinearity-Panos})}: & Q &= \frac{1}{8}
\left[ \psi^3(\zeta+1) + \psi^2(\zeta+1)\psi(\zeta-1) +
\psi(\zeta+1)\psi^2(\zeta-1) \right. \\
& \phantom{Q} &\phantom{=} \left. + \psi^3(\zeta-1) \right].
\end{eqnarray*}
We note that the heteroclinic orbit becomes small as $h \to 0$ under the 
normalization (\ref{scaling}): if $\phi(z) \to \pm 1$ as $z \to \pm
\infty$, then $\psi(\zeta) \to \pm h$ as $\Real\zeta \to \pm
\infty$. The formal asymptotic series (\ref{perturbation-series}) in
the new variables (\ref{scaling}) becomes a new formal series
\begin{equation}
\label{perturbation-series-inner} \hat{\psi}(\zeta) =
\hat{\psi}_0(\zeta) + \sum_{n=1}^{\infty} h^n \hat{\psi}_n(\zeta),
\end{equation}
where each term $\hat{\psi}_n(\zeta)$ can be expanded in
a formal series in descending powers of $\zeta$.
In particular, the leading-order function
$\hat{\psi}_0(\zeta)$ has the general form
\begin{equation}
\label{leading-order-inner} \hat{\psi}_0(\zeta) =
\sum_{n=0}^{\infty} \frac{a_{2n}}{\zeta^{2n+1}}.
\end{equation}
By comparing the series (\ref{perturbation-series-inner}) and
(\ref{leading-order-inner}) with the solutions (\ref{leading-order})
and (\ref{first-order-explicit}) in variables (\ref{scaling}), we
note the correspondence:
$$
a_0 = 2 \sqrt{1 - c^2}, \qquad a_2 = - \frac{(1-c^2)(\gamma +
2\beta) + 6}{9 \sqrt{1 - c^2}}.
$$
We shall be referring to (\ref{perturbation-series-inner}) as the
inner asymptotic expansion. The odd powers of $h$ in the inner
asymptotic expansion (\ref{perturbation-series-inner}) appear due to
the matching conditions with the outer asymptotic expansion
(\ref{perturbation-series}) under the scaling (\ref{scaling}), as
well as due to the non-zero boundary conditions for the heteroclinic
orbits $\psi(\zeta) \to \pm h$ as $\Real\zeta \to \pm \infty$.

\subsection{Leading-order problem for an inner solution}

Convergence of the formal inverse-power series
(\ref{leading-order-inner}) for the leading-order solution
$\hat{\psi}_0(\zeta)$ depends on the values of the Stokes constants
\cite{Tovbis2}. Computation of the Stokes constants is based on
Borel--Laplace transforms of the inner equation (\ref{inner})
\cite{Tovbis1}. Assuming continuity in $h$, we study the
leading-order solution $\psi_0(\zeta) = \lim_{h \to 0} \psi(\zeta)$
of the truncated inner equation
\begin{equation}
\fl c^2 \psi_0''(\zeta) = \psi_0(\zeta+1)-2\psi_0(\zeta) +
\psi_0(\zeta-1) -
Q\left(\psi_0(\zeta-1), \psi_0(\zeta), \psi_0(\zeta+1)\right).
\label{inner-limiting}
\end{equation}
By substituting the series
(\ref{leading-order-inner}) into equation
(\ref{inner-limiting}), one can derive a recurrence relation between the
coefficients in the set $\{ a_n \}_{n=0}^{\infty}$. The Stokes
constants can be computed from the asymptotic behavior of the
coefficients $a_n$ for large $n$. Alternatively, the leading-order
solution $\psi_0(\zeta)$ and the Stokes constants can be defined by
using the Borel--Laplace transform:
\begin{equation}
\label{Laplace} \psi_0(\zeta) = \int_{\gamma} V_0(p) \rme^{-p\zeta} \rmd
p.
\end{equation}
The choice of the contour of integration $\gamma$ determines the
domain of $\psi_0(\zeta)$ in the complex $\zeta$-plane. We define
two solutions $\psi_0^{(s)}(\zeta)$ and $\psi_0^{(u)}(\zeta)$, which
lie on the stable and unstable manifolds respectively, such that
\begin{equation}
\label{stable-unstable} \lim_{\Real\zeta \to +\infty}
\psi_0^{(s)}(\zeta) = 0, \quad \lim_{\Real\zeta \to -\infty}
\psi_0^{(u)}(\zeta) = 0.
\end{equation}
We note that the stable and unstable solutions tend to the 
stationary point at the origin,
since the heteroclinic orbits connect the stationary points
at $\psi = \pm h$ which move to the origin as $h \to 0$.
The three stationary points coalesce to become a
degenerate stationary point at the origin within the truncated
inner equation (\ref{inner-limiting}).

The Borel--Laplace transform (\ref{Laplace}) produces the stable
solution $\psi_0^{(s)}(\zeta)$ when the contour of integration
$\gamma_s$ lies in the first quadrant of the complex $p$-plane
and extends from $p=0$ to $p=\infty$.  Similarly, it produces
the unstable solution $\psi_0^{(u)}(\zeta)$ when the
contour of integration $\gamma_u$ lies in the second quadrant.
We choose the integration contours in such a way that $\arg p \to \pi/2$
as $p \to \infty$, so that the solutions $\psi_0^{(s)}(\zeta)$ and 
$\psi_0^{(u)}(\zeta)$ are defined by (\ref{Laplace}) for all complex $\zeta$
with $\Imag \zeta < 0$.

\begin{figure}
\centering
{\normalsize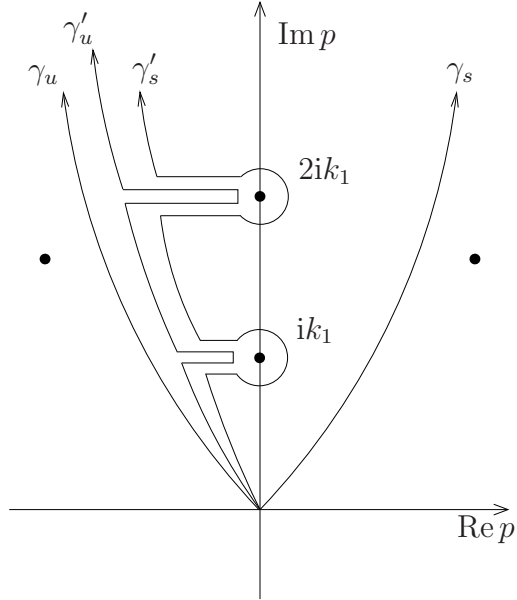}
\caption{\label{pathofintegrationfig}Contours of integration for the
stable and unstable solutions $\psi_0^{(s)}$ and $\psi_0^{(u)}$.  $\gamma_s^{\prime}$
and $\gamma_u^{\prime}$ are 
deformations of the contours $\gamma_s$ and $\gamma_u$ respectively.}
\end{figure}

The Borel transform $V_0(p)$ satisfies the following integral equation,
which follows from (\ref{inner-limiting}) and (\ref{Laplace}):
\begin{equation}
\label{integral-limiting} \left( 4 \sinh^2 \frac{p}{2} - c^2 p^2
\right) V_0(p) = \hat{Q}\left[V_0(p)\right].
\end{equation}
Here, $\hat{Q}\left[V(p)\right]$ denotes a double convolution of $V(p)$
with itself (in this case, the hat is used to denote an operator).
We list below the convolutions $\hat{Q}\left[V(p)\right]$ for
each of the four models under consideration:
\begin{eqnarray*}
\fl \quad \textrm{One-site (\ref{nonlinearity-phi4})} :  &2 \hat{Q} &=
V(p) \ast V(p) \ast V(p); \\
\fl \quad \textrm{Speight-Ward (\ref{nonlinearity-Speight})} :  &18 \hat{Q} &=
\cosh p \; \left[ V(p) \ast V(p) \ast V(p) \right] + 3 \left\{ \cosh
p \; \left[ V(p) \ast V(p) \right] \right\} \\ 
\fl & \ast V&(p) + 3 \left[
\cosh p \; V(p) \right] \ast V(p) \ast V(p) + 2 V(p) \ast V(p) \ast V(p); \\
\fl \quad \textrm{Bender-Tovbis (\ref{nonlinearity-Tovbis})} : &2 \hat{Q} &=
\left[ \cosh p \; V(p) \right] \ast V(p) \ast V(p); \\
\fl \quad \textrm{Kevrekidis (\ref{nonlinearity-Panos})} : \quad &4\hat{Q} &=
\cosh p \; \left[ V(p) \ast V(p) \ast V(p) \right] \\
\fl & & \phantom{=} + \left[ \cosh p
\; V(p) \right] \ast \left[ \rme^p\, V(p) \right] \ast \left[ \rme^{-p}\,V(p)
\right],
\end{eqnarray*}
where the asterisk $\ast$ denotes the convolution integral for the
Borel--Laplace transform:
$$
V(p) \ast W(p) = \int_0^p V(p-p_1) W(p_1) \rmd p_1,
$$
and the integration is performed from the origin to the point $p$ on the
complex plane, along the contour $\gamma$. The
inverse power series (\ref{leading-order-inner}) for the limiting
solution $\psi_0(\zeta)$ becomes the following power series for the
Borel transform $V_0(p)$:
\begin{equation}
\label{Laplace-power} \hat{V}_0(p) = \sum_{n=0}^{\infty} v_{2n}
p^{2n}, \qquad v_{2n} = \frac{a_{2n}}{(2n)!},
\end{equation}
where $v_0 = a_0 = 2 \sqrt{1-c^2}$. The hat denotes a
formal series which might only converge for some values of $p$.
The virtue of the integral form (\ref{integral-limiting})
is that the limiting behavior
of $v_n$ for large $n$ can be related to singularities of $V_0(p)$, which
in turn correspond to the oscillatory tails of $\psi_0(\zeta)$.

If the sliding kink exists, the inverse-power series for
$\psi_0(\zeta)$ will converge for all $\zeta \in \mathbb{C}$ such
that $\Imag \zeta < 0$. This implies that the stable and unstable
solutions $\psi_0^{(s)}(\zeta)$ and $\psi_0^{(u)}(\zeta)$ coincide, i.e.
that the contour $\gamma_s$ in the right half of the complex 
$p$-plane can be continously deformed to the contour $\gamma_u$ in the
left half-plane (see figure \ref{pathofintegrationfig}). If,
however, there are any singularities between the two contours,
then a continuous deformation is possible only if
the residues are zero. The residues
are proportional to the values of the Stokes constants. When
the Stokes constants are nonzero, the formal power series
(\ref{Laplace-power}) for the solution $V_0(p)$ of the integral
equation (\ref{integral-limiting}) diverges for some values
of $p$ in the sector between the contours $\gamma_s$ and $\gamma_u$.

\subsection{Stokes constants}

The Borel transform $V_0(p)$ is singular near the points
in the $p$-plane where the coefficient in front of $V_0(p)$ on 
the left-hand side of the integral equation
(\ref{integral-limiting}) vanishes \cite{Tovbis2}, except for the
point $p = 0$ where the right hand side is also zero.
That is, singularities occur when $(2/p)\sinh(p/2) = \pm c$.
The location of
these singularities is important because the stable and unstable solutions
are not, in fact, uniquely defined by
(\ref{stable-unstable}); different solutions are generated depending on
where the contours lie relative to the singularities of $V_0(p)$ with 
$\Real p \neq 0$.  Exploiting this nonuniqueness, we wish to choose 
the contours $\gamma_s$ and $\gamma_u$ to lie above all the singularities 
with nonzero real part; this will minimise the number of singularities
between the stable and unstable solutions.

It is not difficult to show that the contour $\gamma_s$ extending
from 0 to $\infty$ can be chosen in such a way, i.e. so that that there
are no singularities between it and the imaginary axis.
Indeed, assume, for definiteness, that $c > 0$. 
Let $(2n_c-1)$ be the number of positive roots of the equation
$\sin q = cq$ and denote the real and imaginary
parts of $p/2$ by $\kappa$ and $q$: $p/2=\kappa+iq$. In the $(\kappa,q)$-plane,
consider a rectangular region ${\cal D}$
bounded by the horizontal segments $q=2 \pi n$ and $q=\epsilon$ at the 
top and bottom,
and vertical segments $\kappa=-\epsilon$ and $\kappa=\epsilon$ on 
the left and right.
Here $n$ is any positive integer greater than $n_c$ and $\epsilon>0$
is taken to be small.
Using the argument principle, we can count the
number of (complex) roots of the equation
$\sinh (p/2)=c (p/2)$ in the region ${\cal D}$. We have
\[
\tan \, \arg \frac{p}{2} = \frac{\cosh \kappa \sin q - cq}
{\sinh \kappa  \cos q- c\kappa}.
\label{arg1}
\]
On the right lateral side, where $\kappa = \epsilon$, this becomes
\begin{equation}
\tan \, \arg \frac{p}{2} \approx \frac{1}{\epsilon} \frac{ \sin q - cq}{ \cos q- c }.
\label{arg2}
\end{equation}
As we move from $q=\epsilon$ to $q= 2 \pi n$, the  numerator
in  (\ref{arg2}) will change sign $(2n_c-1)$ times. In a similar way, 
moving down along
the left side there will be $(2n_c-1)$ more zero crossings, while no
zero crossings will occur along the  horizontal segments. This
means that the argument can change by no more than $(4n_c-2) \pi$ and
hence there are
at most $(2n_c-1)$ roots in the region ${\cal D}$, no matter how large
$n$ is. Similarly, we can show that the equation 
$\sinh (p/2)=-c (p/2)$ has no more than $2n_c$ roots in the region ${\cal D}$,
if $2n_c$ is the number of positive roots of $\sin q= -cq$.
The upshot is that for any finite $c$,
there are only a finite number of singularities with small real parts;
the singularities cannot accumulate to the imaginary axis.
For $c \neq 0$, the singularities
with nonzero real parts lie on the curves
\[
\fl q = \pm \sqrt{\frac{1}{c^2}\cosh^2\kappa -
\kappa^2\coth^2\kappa} \to \pm\frac{1}{c}\cosh\kappa 
\quad \textrm{as $|\kappa|\to\infty$}.
\]
Accordingly, in order for the integration contours $\gamma_s$ and
$\gamma_u$ to lie above these singularities, they must be curvilinear 
(and not just rays) as shown in figure \ref{pathofintegrationfig}.

Having chosen the contours $\gamma_s$ and $\gamma_u$ to lie above the 
singularities in the first and second quadrants respectively, the
only singularities of $V_0(p)$ that determine whether the stable
solution $\psi_0^{(s)}(\zeta)$ can be continuously transformed into 
the unstable solution $\psi_0^{(u)}(\zeta)$ are those at non-zero 
pure imaginary values of $p$. We will be referring to these values
as resonances. The set of resonances ${\cal R}_c$ is defined by the
transcendental equation
\begin{equation}
\label{resonances} {\cal R}_c = \left\{ p = ik, k \in \mathbb{R}_+ :  \quad
\frac{2}{k} \; \sin \frac{k}{2} = \pm c \right\}.
\end{equation}
When $c = 0$, the set ${\cal R}_0$ is infinite-dimensional and can be
described explicitly:
\[
{\cal R}_0 = \left\{ p = 2 \pi n \rmi, \quad n \in \mathbb{N} \right\}.
\]
Let $p_1 = \rmi k_1$ be the smallest imaginary root in the set ${\cal R}_c$. It
is clear from (\ref{resonances}) that $0 < k_1 < 2 \pi$ for $c \in (0, 1)$,
so that $k_1 \to 2 \pi$ as $c \to 0^+$ and $k_1 \to 0$ as $c
\to 1^-$. The set of resonances ${\cal R}_c$ is finite-dimensional for 
$c \in (0, 1)$ and it consists of only one root $p_1 = \rmi k_1$ for 
$c \in (c_1, 1)$,
where $c_1 \approx 0.22$.

Due to the resonances, a function $\psi_0(\zeta)$ that satisfies
the truncated inner equation (\ref{inner-limiting}) may have
oscillatory tails as $|\Real\zeta| \to \infty$. Adding the
solutions of equation (\ref{inner-limiting}) linearised about
$\hat{\psi}_0(\zeta)$,
the general bounded solution of (\ref{inner-limiting}) 
in the limit $|\zeta| \to \infty$ can be represented as \cite{Tovbis1}:
\begin{eqnarray}
\label{wave-decomposition} \psi_0(\zeta) = \hat{\psi}_0(\zeta) +
\sum_{\rmi k_n \in {\cal R}_c} \alpha_n \hat{\varphi}_n(\zeta) \rme^{-\rmi k_n
\zeta} + \textrm{multiple harmonics}.
\end{eqnarray}
Here, $\hat{\psi}_0(\zeta)$ is given by the power series
(\ref{leading-order-inner}); $\alpha_n$ are coefficients which we will
be referring to as amplitudes in what follows; $k_n
> 0$ are roots of (\ref{resonances}) for $p = \rmi k_n$, 
and the functions $\hat{\varphi}_n(\zeta) \rme^{-\rmi k_n \zeta}$,
$n \geq 1$, satisfy the linearised truncated inner equation
(\ref{inner-limiting}). In particular, the equation for the
leading-order term ${\hat \varphi}_1(\zeta)$ is
\begin{eqnarray}
\fl\rme^{-\rmi k_1}  \hat{\varphi}_1(\zeta+1) + (c^2 k_1^2 - 2)
\hat{\varphi}_1(\zeta) + \rme^{\rmi k_1} \hat{\varphi}_1(\zeta-1) + 2 \rmi
c^2 k_1 \hat{\varphi}_1'(\zeta) - c^2 \hat{\varphi}_1''(\zeta) \nonumber\\
 = D_1 Q \, \hat{\varphi}_1(\zeta -1) +  D_2 Q \,
\hat{\varphi}_1(\zeta) + D_3 Q \, \hat{\varphi}_1(\zeta +1),
\label{inner-limiting-first-order}
\end{eqnarray}
where $D_{1,2,3}Q$ are the partial derivatives of
$Q(\psi_0(\zeta-1),\psi_0(\zeta),\psi_0(\zeta+1))$ with respect to its first,
second and third argument respectively, evaluated at $\psi_0 =
\hat{\psi}_0(\zeta)$.

If the amplitude $\alpha_n$ is nonzero for some $n$, the formal 
power series (\ref{leading-order-inner}) does not
converge because the solution (\ref{wave-decomposition})
does not decay as $|\Real\zeta| \to
\infty$. The amplitudes $\alpha_n$ are proportional to the Stokes
constants computed for the formal power series
(\ref{leading-order-inner}). 
Each oscillatory term in the sum (\ref{wave-decomposition}) becomes
exponentially small in $h$ when we transform from $\zeta$ to $z$ 
using the transformation (\ref{scaling}). 
Since $p_1 = \rmi k_1$ is the element of ${\cal R}_c$ with the smallest 
imaginary part, it follows that the $n = 1$ term dominates the sum in 
(\ref{wave-decomposition}) when the transformation (\ref{scaling}) 
is made (unless $\alpha_1 = 0$). Furthermore, when $c \in (c_1, 1)$, 
where $c_1 \approx 0.22$, it is the \emph{only} term in the sum 
since the resonant
set ${\cal R}_c$ consists of just the one root $p_1 = \rmi k_1$.
We shall, therefore, only be concerned with
the leading-order Stokes constant, which multiplies the function
$\hat{\varphi}_1(\zeta)$.

If $\hat{\psi}_0(\zeta)$ is given by the power series
(\ref{leading-order-inner}), the solution of the linearized
equation (\ref{inner-limiting-first-order}) can also be represented by
a formal power series:
\begin{equation}
\label{power-series-phi} \hat{\varphi}_1(\zeta) = \zeta^r
\sum_{\ell=0}^{\infty} b_{\ell} \zeta^{-\ell},
\end{equation}
where we can set $b_0 = 1$ due to the linearity of
(\ref{inner-limiting-first-order}).
Substituting (\ref{leading-order-inner}) and
(\ref{power-series-phi}) into (\ref{inner-limiting-first-order}) and
using (\ref{resonances}), the coefficient $b_1$ can be determined from
\begin{eqnarray*}
\fl 2 \rmi r \zeta^{r-1} \left( c^2 k_1 - \sin k_1 \right)
+ \zeta^{r-2} \big[ r(r-1)(\cos k_1 - c^2) \\
+ 2 \rmi b_1 (r-1) (c^2 k_1 -
\sin k_1) - 6(1 - c^2) \big] + \order(\zeta^{r-3}) = 0.
\end{eqnarray*}
In this equation, the coefficient of each power of $\zeta$ should be set
to zero.  In order to set the coefficent in front of the first term
to zero in the situation where $c \neq 0$, we must choose $r = 0$.
The second term then gives
$$
b_1 = \frac{3 \rmi (1-c^2)}{c^2 k_1 - \sin k_1},
$$
after which all the other coefficients $b_2$, $b_3$, \ldots, can be
computed recursively.  On the other hand, in the situation with
$c = 0$, we have $k_1 = 2\pi$ and the coefficient in front 
of $\zeta^{r-1}$
is zero regardless of the value of $r$.
Setting the coefficient in front of $\zeta^{r-2}$ to
zero requires that we choose either $r = 3$ or $r = -2$, and hence
we have two different descending-power series, one starting with 
$\zeta^3$ and the other one with $\zeta^{-2}$.  We shall focus on the
former as it dominates the latter in the limit $\zeta \to \infty$.
Again, the succeeding terms in (\ref{power-series-phi}) are determined 
recursively. 

Thus, we have established that the leading-order oscillatory term in the 
expansion (\ref{wave-decomposition}) behaves as 
\begin{equation}
\cases{
\alpha_1 \left[1 + \frac{b_1}{\zeta} + \order{\left(\frac{1}{\zeta^2}\right)}\right]\rme^{- \rmi k_1\zeta} & 
for $c \neq 0$ and \\
\alpha_1\left[\zeta^3 + b_1\zeta^2 + \order{\left(\zeta^1\right)}\right]\rme^{-2\pi\rmi\zeta} &
for $c = 0$.} 
\label{tails}
\end{equation}
For $c \neq 0$, the two leading order terms in the expression above are
generated by, respectively, a simple pole and a logarithmic singularity 
of the Borel transform $V_0(p)$ at $p = \rmi k_1$.  For $c = 0$ they are
generated by a quadruple pole of $V_0(p)$ at $p = 2 \pi \rmi$. From the
fact that $V_0$ is an even function of $p$, we deduce the
structure of this function near the poles:
\begin{equation}
V_0(p) \to
\cases{
\frac{k_1^2K_1(c)}{p^2 + k_1^2} - \sigma(c)\ln(p^2+k_1^2) + \ldots & (for $c\neq 0$) \\
\frac{6(2\pi)^8S_1}{\left(p^2 + 4 \pi^2\right)^4} + \frac{2(2\pi)^6\rho}{\left(p^2 + 4 \pi^2\right)^3} + \ldots & (for $c = 0$)}
\label{poles}
\end{equation}
as $p \to \pm \rmi k_1$. Here $K_1(c)$ and $S_1$ are the leading-order
Stokes constants for $c \neq 0$ and $c = 0$, respectively; $\sigma(c)$
and $\rho$ are independent of $p$, and $\ldots$ stands for terms with
even slower growth as $p\to ik_1$.

To show that these singularities do indeed give rise to the oscillatory 
tails in (\ref{tails}), we compare the two integrals $\psi_0^{(s)}$
and $\psi_0^{(u)}$ for a given value of $\zeta$.  To this end, we deform 
the paths of integration $\gamma_s$ and $\gamma_u$ to $\gamma_s^{\prime}$
and $\gamma_u^{\prime}$ respectively, without crossing any singularities.
This is illustrated in figure \ref{pathofintegrationfig}.
There are two contributions to the difference $\psi_0^{(s)}(\zeta) 
- \psi_0^{(u)}(\zeta)$. The first comes from integrating around the pole
at $p = \rmi k_1$, and is equal to $2\pi \rmi$ times the residue of the 
function $V_0(p)\rme^{-p\zeta}$ at $p = \rmi k_1$, determined from (\ref{poles}).
The second contribution (manifest only in the $c \neq 0$ case) arises
because the \emph{integrand} increases as the singularity is
encircled, since it is a branch point of the logarithm.  Since $\ln z$
can be written as $\ln|z| + \rmi\arg z$, where $z = p-\rmi k_1$, 
we see that $V_0(p)$ increases by
$-2\pi\rmi\sigma(c)$ as the branch point $p = \rmi k_1$ is encircled in
the $c \neq 0$ case. Therefore, the difference in the integrand of 
(\ref{Laplace}) along the paths $\gamma_s^{\prime}$ and 
$\gamma_u^{\prime}$ is $-2\pi\rmi\sigma(c)\rme^{-p\zeta}$,
which must be integrated along the
path of integration from $p = \rmi k_1$ to infinity, to give 
$-2\pi\rmi\sigma(c)\rme^{-\rmi k_1\zeta}/\zeta$.
(We have considered the integration on a Riemann
surface in order to account for branch points.)
Adding together the two contributions discussed above, we have
\begin{equation}
\fl\psi_0^{(s)}(\zeta)-\psi_0^{(u)}(\zeta) = \cases{
\left[ \pi k_1K_1(c) - \frac{2\pi\rmi\sigma(c)}{\zeta} + 
\order\left(\frac{1}{\zeta^2}\right) \right] \rme^{-\rmi k_1\zeta} & for $c \neq 0$\\
-\frac{1}{128}[16 \pi^3 \rmi S_1 \zeta^3 \\
\qquad + (192\pi^4S_1+\rho)\zeta^2 + \order(\zeta^1) ]
\rme^{-2\pi\rmi\zeta} & for $c = 0$.}
\label{stable-minus-unstable}
\end{equation}
If we take the limit $\Real \zeta \rightarrow -\infty$, then the
unstable solution $\psi_0^{(u)}(\zeta)$ decays to zero 
as a power law, according to the expansion
(\ref{leading-order-inner}). Thus, 
the stable solution $\psi_0^{(s)}(\zeta)$
has the oscillatory tail given by the representation
(\ref{wave-decomposition}) with the amplitude factor
\begin{equation}
\label{alpha-1-expression} \alpha_1 =
\cases{
\pi k_1 K_1(c) & for $c \neq 0$ \\
-\frac{\rmi\pi^3}{8} S_1 & for $c = 0$.}
\end{equation}
Similarly, if we take the limit $\Real\zeta \to +\infty$, then
the stable solution $\psi_0^{(s)}(\zeta)$ decays to zero, while the
unstable solution $\psi_0^{(u)}(\zeta)$ has the representation
(\ref{wave-decomposition}) with the amplitude factor given
by the negative of expression (\ref{alpha-1-expression}).  By comparing
the other terms on the right-hand side of (\ref{stable-minus-unstable})
to the corresponding terms in (\ref{tails}), $\sigma(c)$ and $\rho$ 
can be uniquely determined.

We now match the leading-order singular behaviour of $V_0(p)$ near 
$p = \pm \rmi k_1$, given by (\ref{poles}), to
the formal power series (\ref{Laplace-power}). Expanding the
expressions in (\ref{poles}) as power series gives us
\begin{equation}
\fl V_0(p) \to
\cases{
K_1(c) - \sigma(c)\ln(k_1^2) + \ldots \\
\qquad + \sum_{n=1}^{\infty} (-1)^n k_1^{-2n} \left(K_1(c)+\frac{\sigma(c)}{n} + \ldots\right) p^{2n}
& for $c\neq 0$ \\
\sum_{n=0}^{\infty} \frac{(-1)^n (n+2)(n+1)}{(2\pi)^{2n}}\left[(n+3)S_1 + \rho + \ldots\right] p^{2n} & for $c = 0$,}
\label{poles-expanded}
\end{equation}
as $p \to \pm \rmi k_1$. These series converge for all $|p| < k_1$;
in particular, they are valid for $p \to \pm \rmi k_1$, provided $|p| < k_1$.
Hence we can replace (\ref{poles}) with (\ref{poles-expanded}) in this neighbourhood.
In (\ref{poles-expanded}), the ellipses stand for
coefficients of the expansion of terms with a slower growth as 
$p\to \pm\rmi k_1$
which were dropped in (\ref{poles}).  The discarded terms
would modify the coefficients of the power series (\ref{poles-expanded});
however, there are terms which would not be affected by these modifications,
namely terms with large $n$.  For example, the coefficients
proportional to $\sigma(c)$ and $\rho$ in (\ref{poles-expanded}) are a factor
of $n$ smaller than those proportional to $K_1(c)$ and $S_1$; the discarded
coefficients would be even smaller.  Therefore the leading singular
behaviour of $V_0(p)$ as $p \to \pm \rmi k_1$ is determined just by
the large-$n$ coefficients of the power series (\ref{poles-expanded}), and
hence only the large-$n$ coefficients should be matched to the coefficients
of the expansion (\ref{Laplace-power}).
This gives the Stokes constant as a limit of
the coefficients $v_{2n}$ of the series (\ref{Laplace-power}):
\begin{equation}
\label{Stokes}
\cases{
K_1(c) = \displaystyle \lim_{n \to \infty} (-1)^n k_1^{2n} v_{2n} & for
$c \neq 0$ \\
S_1 = \displaystyle \lim_{n \to \infty} \frac{(-1)^n (2 \pi)^{2n}
v_{2n}}{(n+3)(n+2)(n+1)} & for $c = 0$.}
\end{equation}
This  formula is used in the next section for
numerical computations of the leading-order Stokes constant $K_1(c)$
for $c \neq 0$.

Note that, since (\ref{Laplace-power}) matches (\ref{poles-expanded})
in the limit $n\to\infty$, the formal power series $\hat{V}_0(p)$ 
also has radius of convergence $k_1$.  However, the formal
inverse-power series $\hat{\psi}_0(\zeta)$ diverges for all $\zeta$ unless
$\hat{V}_0(p)$ converges everywhere (which requires that all the Stokes 
constants be zero).

Next, we note that as $c \to 0$, the Stokes constant $K_1(c)$ does not 
tend to $S_1$, its value at $c = 0$. This discontinuity is due to the fact 
that, as $c \to 0$, pairs of simple roots in the resonant set ${\cal R}_c$ 
coalesce. (E.g. $\rmi k_1$ coalesces with $\rmi k_2$
at $2 \pi \rmi$, and so on.)  As a result, all 
roots are double and the representation of $\hat{\varphi}_1(\zeta)$ 
is discontinuous at $c = 0$, with the power degree
$r$ of the prefactor in (\ref{power-series-phi}) jumping from $r = 0$ for $c \neq 0$ 
to $r = 3$ for $c = 0$. 
In particular, in exceptional models, i.e., discrete models with continuous families
of stationary kinks (like (\ref{nonlinearity-Speight}), 
(\ref{nonlinearity-Tovbis}) and
(\ref{nonlinearity-Panos})) the constant
$S_1$ is \emph{a priori} zero while the limit of $K_1(c)$ as $c\to 0$
may be nonvanishing. In fact, numerical computations
of the top limit in (\ref{Stokes}) indicate that the Stokes constant
blows up as $c\to 0$.  Renormalisation of $K_1(c)$ for small $c$
is, however, a nontrivial asymptotic problem which is beyond the
scope of our current investigation.

For $c \in (c_1, 1)$, where $c_1 \approx 0.22$, the resonant set 
${\cal R}_c$ contains only one root $\rmi k_1$ and, therefore, there
is just one Stokes constant $K_1(c)$, which completely determines
the convergence of the formal power
series for $\hat{\psi}_0(\zeta)$. If $K_1(c_0) = 0$ at some 
point $c_0  \in (c_1,1)$, the stable and unstable solutions
$\psi_0^{(s)}(\zeta)$ and $\psi_0^{(u)}(\zeta)$ coincide to leading
order. Arguments based on the implicit function theorem (see
\cite{TovPel}) reveal a heteroclinic bifurcation
which occurs on crossing a smooth curve $c=c_*(h)$ 
on the $(c,h)$-plane,
with $c_*(0)=c_0$.

On the other hand, for $c \in (0, c_1)$ the resonant set 
${\cal R}_c$ contains
more than one root. 
If $K_1(c_0)=0$ for some $c_0  \in (0,c_1)$, this alone is not
sufficient for the convergence of the formal power series
${\hat \psi}_0(\zeta)$.  The higher-order Stokes constants
$K_2(c)$, $K_3(c)$, \ldots, must be introduced and computed from the
asymptotic behavior of the power series $\hat{V}_0(p)$.

As we shall show in the next section, the function $K_1(c)$
does have zeros in the case of the discretizations 
(\ref{nonlinearity-Speight}) and (\ref{nonlinearity-Panos}).
All these zeroes are ``safe''; that is, all $c_0$ values
lie in the interval $(c_1,1)$, so that the higher-order Stokes constants 
do not have to be computed.

\section{Numerical computations of the Stokes constant}

\label{section3}

In this section, we report on the numerical computation of the 
Stokes constants $K_1(c)$ for the four different discretizations 
of the $\phi^4$ model (\ref{phi-4}) under consideration. Our
numerical method utilises the expression (\ref{Stokes}) of the Stokes
constant in terms of the coefficients of the formal power series
solution (\ref{Laplace-power}). First, we obtain the recurrence
relation for the coefficients in the set $\{ v_n \}_{n=0}^{\infty}$
by substituting the power series expansion (\ref{Laplace-power})
into the limiting integral equation (\ref{integral-limiting}), and
using the convolution formula
\begin{equation}
\label{convolution-powers} p^n \ast p^m = \frac{n! \;
m!}{(n+m+1)!} p^{n+m+1}.
\end{equation}
After that, we compute the asymptotic behavior of these coefficients
as $n\to\infty$ and
evaluate the limit (\ref{Stokes}) numerically for a fixed value of
$c \neq 0$.

In order to calculate the Stokes constant for the
four models in a uniform way, we write a general
symmetric homogeneous cubic polynomial $Q(u_{n-1},u_n,u_{n+1})$ as
\begin{equation}
\label{cubic-polynomial}
Q = \sum_{\alpha = -1}^{1}\sum_{\beta =
\alpha}^{1}\sum_{\gamma = \beta}^{1}a_{\alpha,\beta,\gamma}\,
u_{n+\alpha}u_{n+\beta}u_{n+\gamma}, 
\end{equation}
where $a_{\alpha,\beta,\gamma}$ are numerical coefficients, with $\alpha,\beta,\gamma \in
\{-1,0,1\}$ and
$\alpha \le \beta \le \gamma$.  The symmetry implies that
$a_{\alpha,\beta,\gamma} = a_{-\gamma,-\beta,-\alpha}$ and therefore
it is sufficient to specify just six coefficients.
The values of these coefficients for the four
nonlinearities in question are given in Table \ref{coeffs}.

\begin{table}[tbh]
\centering%
\begin{tabular}{|l|cccccc|}
\hline
Model & \; $a_{1,1,1}$ \; & \; $a_{0,0,0}$ \; & \; $a_{0,1,1}$ \; & \;
$a_{0,0,1}$ \; & $a_{-1,1,1}$ \; & \; $a_{-1,0,1}$ \; \\
\hline
One-site (\ref{nonlinearity-phi4}) & 0 & $1/2$ & 0 & 0 & 0 & 0 \\
\hline
Speight-Ward (\ref{nonlinearity-Speight}) & $1/36$ &
$1/9$ & $1/12$ & $1/12$ & 0 & 0 \\
\hline
Bender-Tovbis (\ref{nonlinearity-Tovbis}) & 0 & 0 & 0 & $1/4$ & 0 & 0 \\
\hline
Kevrekidis (\ref{nonlinearity-Panos}) & $1/8$ & 0 & 0 & 0 & $1/8$ & 0 \\
\hline
\end{tabular}
\caption{\label{coeffs}The coefficients
$a_{\alpha,\beta,\gamma}=a_{-\gamma,-\beta,-\alpha}$ of the cubic
polynomial (\ref{cubic-polynomial}) for the four models under consideration.}
\end{table}

By applying the Borel--Laplace transform (\ref{Laplace}) to equation
(\ref{inner-limiting}) with $Q$ as in (\ref{cubic-polynomial}),
we obtain the corresponding cubic convolution
function $\hat{Q}[V(p)]$ on the right hand side of the integral equation
(\ref{integral-limiting}):
\begin{equation}
\label{integraleq} \hat{Q}\left[V(p)\right] = \sum_{\alpha = -1}^{1}\sum_{\beta =
\alpha}^{1}\sum_{\gamma = \beta}^{1}a_{\alpha,\beta,\gamma} \;
\rme^{\alpha p} V(p) \ast \rme^{\beta p}V(p) \ast \rme^{\gamma p}V(p).
\end{equation}
To derive the recurrence formula for the coefficents $v_{2n}$ in 
(\ref{Laplace-power}), it will be more convenient to consider
the power series expansion which consists of both even and odd powers of $p$:
\begin{equation}
\label{general-Laplace-power} \hat{V}_0(p) = \sum_{n=0}^{\infty} v_n p^n.
\end{equation}
We now substitute the series (\ref{general-Laplace-power}) into
(\ref{integral-limiting}) with $\hat{Q}\left[V(p)\right]$ given by
(\ref{integraleq}) and use the convolution formula
(\ref{convolution-powers}).  Equating the coefficients of $p^{n+2}$ 
where $n = 0,1,2,\ldots$, in the resulting equation, we find that
\begin{eqnarray}
\label{recurrence}
\fl  \sum_{i = 0}^{[n/2]} \frac{2}{(2i + 2)!} v_{n-2i} - c^2 v_n =
\sum_{\alpha = -1}^{1}\sum_{\beta = \alpha}^{1}\sum_{\gamma = \beta}^{1}
\frac{a_{\alpha,\beta,\gamma}}{(n+2)(n+1)}\Bigg\{
\sum_{i=0}^n\left(\sum_{k=0}^{n-i}\frac{\alpha^k}{k!}v_{n-i-k}\right) \nonumber\\
\times \left[\sum_{j=0}^i\left(\sum_{l=0}^j\frac{\beta^l}{l!}v_{j-l}\right)
\left(\sum_{m=0}^{i-j}\frac{\gamma^m}{m!}v_{i-j-m}\right)\frac{j!(i-j)!}{i!}\right]\frac{i!(n-i)!}{n!}\Bigg\},
\end{eqnarray}
where $[n/2]$ is the integer part of $n/2$ and $0^0=1$. Equation
(\ref{recurrence}) is a recurrence relation between the coefficients
$\{ v_n \}_{n=0}^{\infty}$. Solving equation (\ref{recurrence}) with $n = 0$,
we get $v_0 = 2 \sqrt{1-c^2}$. Note that this result is independent
of the choice of $a_{\alpha, \beta, \gamma}$, i.e. independent
of the model.
Letting $v_1 = 0$ and making use of the symmetry of $Q$, one
can show by induction that the coefficients of all odd powers in
(\ref{general-Laplace-power}) are zero (as we concluded
previously on the basis that the outer expansion is odd).

To prevent overflow or underflow when
evaluating the recurrence relation numerically, we shall work with
the normalised coefficients
\begin{equation*}
w_n = (-1)^n k_1^{2n} v_{2n}
\end{equation*}
so that the Stokes constant (\ref{Stokes}) for $c \neq 0$ is given by
\begin{equation}
\label{numerical-limit} K_1(c) = \lim_{n\rightarrow \infty} w_n.
\end{equation}
Reformulating (\ref{recurrence}) in terms of $w_n$, we
use the relation (\ref{numerical-limit}) to compute $w_n$ numerically. 
We truncate the
sums involving $1/(2i+2)!$, $1/l!$ and $1/m!$ when these factors 
become smaller than $10^{-50}$, and evaluate the sums involving the
combinatorial factors in two halves. In the first, the summation
index increases from zero to the halfway point, and in the second
it decreases from its maximum.  This ensures that the
combinatorial factors are always decreasing from one step to the
next so that they can be accurately determined recursively.  We
also truncate these sums when the combinatorial factors fall below
$10^{-50}$.

\begin{figure}[tbp]
\mbox{\footnotesize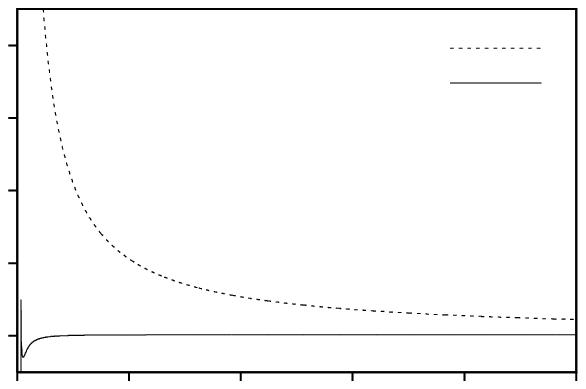}\quad
\mbox{\footnotesize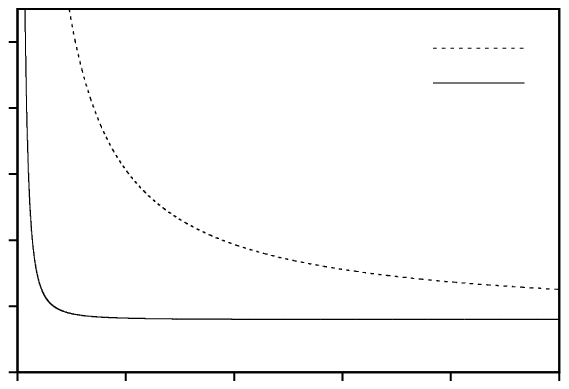}
\caption{\label{sequencegraphs}Convergence of the sequence
$w_n$ and its accelerated counterpart ${\tilde w}_n$ to the Stokes constant
$K_1(c)$. The dashed line depicts the values of $w_n$ and the solid line marks 
the accelerated sequence ${\tilde w}_n$.  Left: the Kevrekidis
discretization (\ref{nonlinearity-Panos}) with $c = 0.5$. Right:
Speight-Ward discretization (\ref{nonlinearity-Speight}) with $c =
0.005$.}
\end{figure}

These expedients result in a numerical routine fast enough to allow
for evaluation of
the recurrence relation up to very large $n$; this is essential
given the slow convergence of $w_n$ to a constant. Matching 
(\ref{Laplace-power}) to (\ref{poles-expanded}) yields
\[
v_{2n} \to (-1)^n k_1^{-2n}[K_1(c) + \sigma(c)/n] \quad \textrm{as $n \to \infty$};
\]
therefore,
the rate at which $w_n$ converges to $K_1(c)$ is of order $1/n$:
\begin{equation}
  w_n=K_1(c) +\frac{\sigma(c)}{n} + \frac{\tilde{\sigma}(c)}{n^2}+
                            {\cal O}\left( \frac{1}{n^3} \right).
\label{convergence-of-w_n}
\end{equation}
Although the convergence of
$w_n$ to $K_1(c)$ is extremely slow, we can accelerate the
process by using (\ref{convergence-of-w_n}). Defining
\[
             {\tilde w}_n \equiv w_n+ n(w_n-w_{n-1}),
\]
we get
\[
           {\tilde w}_n =K_1(c) -\frac{\sigma(c)+\tilde{\sigma}(c)}{n^2}
      + {\cal O} \left( \frac{1}{n^3} \right).
\]
The convergence of the sequence ${\tilde w}_n$ is much faster than that
of $w_n$; see Figure \ref{sequencegraphs}. The relative error
\[
         E(n)= \frac{\sigma(c)+\tilde{\sigma}(c)}{ n^2} \frac{1}{K_1(c)}
\]
can be written as
\[
         E(n) = \frac{n}{2} \frac{{\tilde w}_n-{\tilde w}_{n-1}}{{\tilde w}_n}   
\]                                                                
plus terms of order $1/n^4$. This gives an empirical criterion for the
termination of the process. We continued our computations until $E(n)$
reached a value smaller than $10^{-3}$, i.e. until the percentage error
dropped below 0.1\%.  For $c>0.5$, the value of $n$ to which
we have to compute in order to achieve
this accuracy is less than $100$,
increasing for smaller values of $c$ to approximately $5\,000$ for $c = 0.005$. 
Consequently, the above numerical algorithm is not suited
to the study of the $c \to 0$ limit, and would have to be modified
for that purpose.

\begin{figure}[tbp]
\mbox{\footnotesize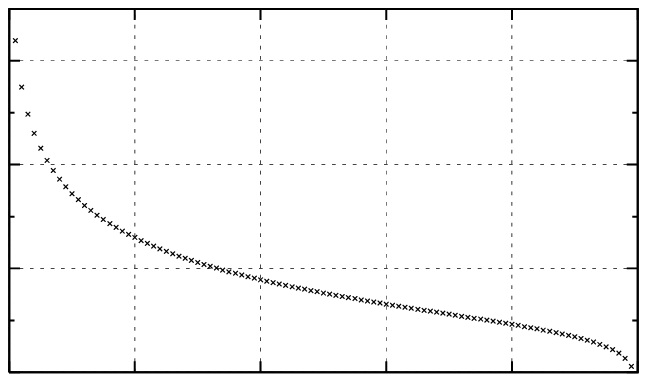}\quad
\mbox{\footnotesize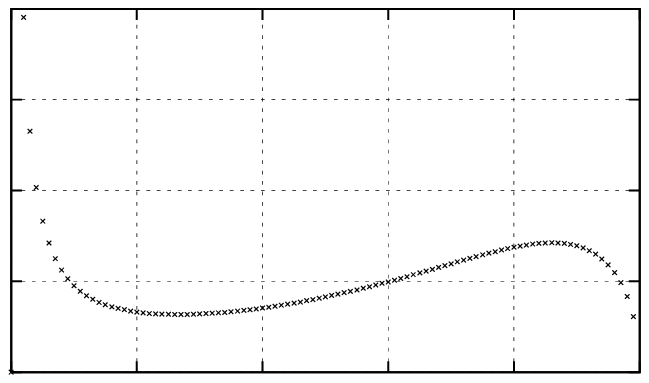}
\par\smallskip\smallskip\par
\mbox{\footnotesize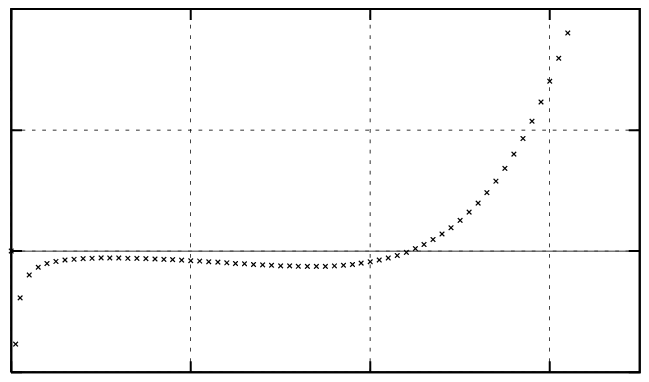}\quad
\mbox{\footnotesize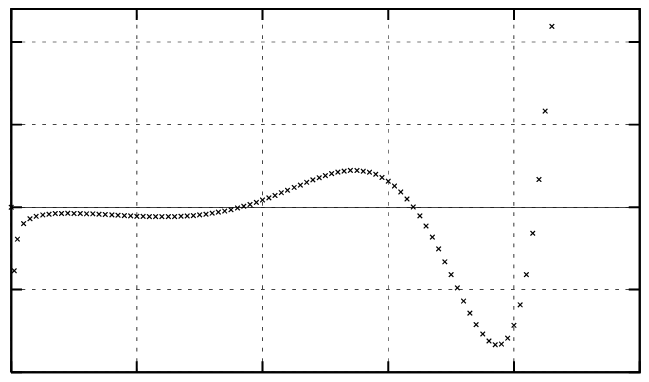}
\caption{\label{stokesgraphs}The Stokes constant $K_1$ as a function
of the kink's speed $c$ for the four discretizations of the $\phi^4$ model.
Clockwise from the top
left: one-site (\ref{nonlinearity-phi4}); Bender-Tovbis
(\ref{nonlinearity-Tovbis}); Kevrekidis (\ref{nonlinearity-Panos});
Speight-Ward (\ref{nonlinearity-Speight})}
\end{figure}

Figure \ref{stokesgraphs} displays the Stokes constant computed
using the above numerical procedure, for the four models of
Table \ref{coeffs}.  We see that the Stokes
constant $K_1(c)$ vanishes almost nowhere in
$c \neq 0$ in all four models. There are, however, several
isolated zeros: $K_1(c) = 0$ for $c_0 \approx 0.45$ in the case of the
Speight-Ward nonlinearity (\ref{nonlinearity-Speight}) and for 
$c_0  \approx 0.37$, $0.63$ and $0.83$ in the case
of the Kevrekidis discretization (\ref{nonlinearity-Panos}). 
Importantly, all of
these lie in the region $(c_1, 1)$ where
the resonance set (\ref{resonances}) consists of only one
value, $p_1 = \rmi k_1$. (Here $c_1 \approx 0.22$.)  
Therefore, there is a sliding kink in the $h \to 0$ limit for each
of these isolated values of velocity.  Furthermore, strong parallels between
our current setting and that of solitons of the fifth-order KdV equation
\cite{TovPel} suggest that sliding kinks should exist along
a curve on the $(c,h)$ plane emanating from each of the points
$(c_0, 0)$.  In other words, we conjecture that there is a radiationless
kink travelling
with a certain particular speed $c_*(h)$ for each $h$ in the case of the
Speight-Ward nonlinearity, and that there are three such
velocities (for each $h$) in the case of the Kevrekidis model.
For small $h$, $c_*(h)$ should be close to the above values $c_0$. 

In order to verify the existence of kinks sliding at these isolated
velocities by an independent method, we solved the differential
advance-delay equation (\ref{advance-delay}) numerically. The infinite
line was approximated by an interval of length
$2L = 200$, with the antiperiodic boundary conditions $\phi(-L) =
-\phi(L)$.  We utilised Newton's iteration with an eighth-order
finite-difference approximation of the second derivative; the
step size was chosen to be $h/10$.
The continuum solution (\ref{travelling}) was used as an initial
guess.

\begin{figure}[tbp]
\mbox{\footnotesize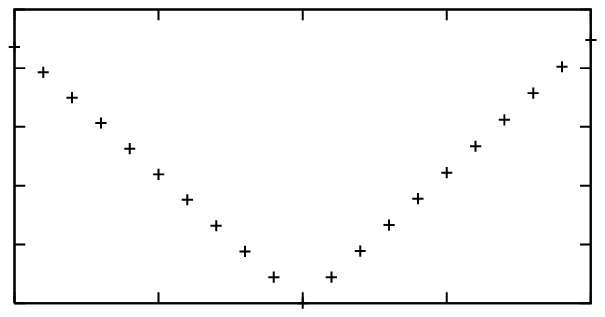}\quad
\mbox{\footnotesize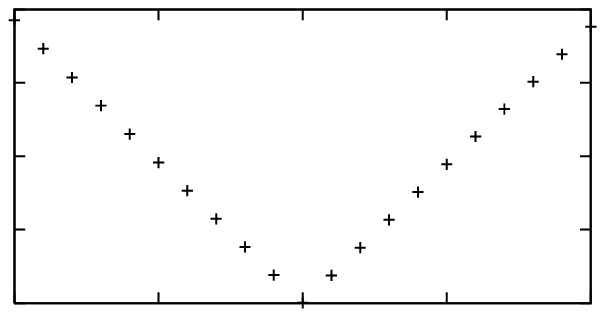}
\par\smallskip\smallskip\par
\mbox{\footnotesize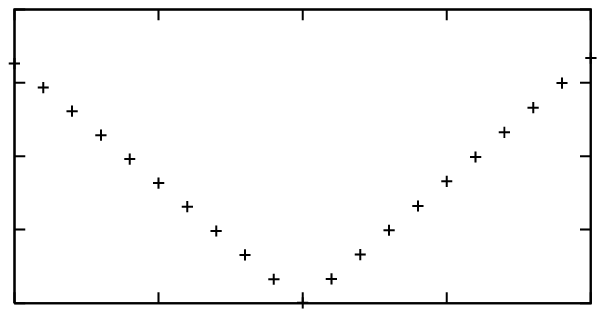}\quad
\mbox{\footnotesize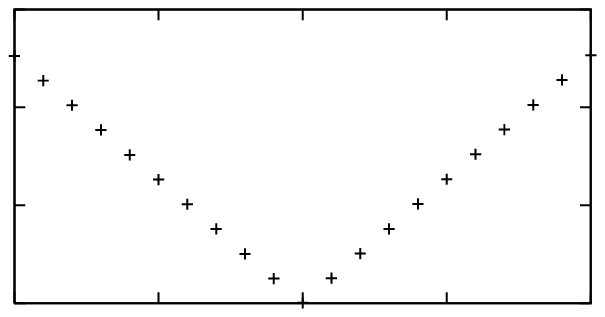}
\caption{\label{radiationzero} The numerical evidence for the
disappearance of radiation at isolated values of $c = c_*$ in
Speight and Ward's model (top left panel) and Kevrekidis' model (three other panels).
The magnitude of the oscillatory tails, as defined in the text,
is plotted as a function of $c$ for a fixed value of $h$.
The minimum radiation is attained at the value $c_*(h)$ which
is found numerically.  (This $c_*(h)$ is of course slightly different
from the value $c_*(0)$ for which the Stokes constant vanishes.)}
\end{figure}

If we find a solution to the advance-delay equation with $\phi(z)$ 
decaying to a constant for large positive and negative $z$, then we
regard this solution as (a numerical approximation to) a radiationless
travelling kink. We were able to tune $c$ for a fixed value of $h$ so 
that the radiation was reduced to the order of $10^{-12}$, whereupon the
finite accuracy of our numerical scheme prevented any further
reduction.  To make sure that the radiation does vanish rather
than reaching a local minimum but remaining nonzero, we plot the 
average magnitude of the radiation
near the ends of the interval as a function of $c$, for fixed $h$.
This is defined as
the average of $\left[\phi(z)-\overline{\phi}\right]^2$ over the
last 20 units of the interval, where $\overline{\phi}$ is the average 
value of $\phi(z)$ over these last 20 units.
The results are shown
in figure \ref{radiationzero}.  Note the straight-line behavior of the
graphs near the isolated values of $c$; this indicates that the
coefficient of the sinusoid superimposed over the kink's flat asymptote
crosses through zero (rather than attaining a small but nonzero minimum).
The supression of radiation at the isolated points is thereby verified.

\begin{figure}[tbp]
\centering%
\mbox{\footnotesize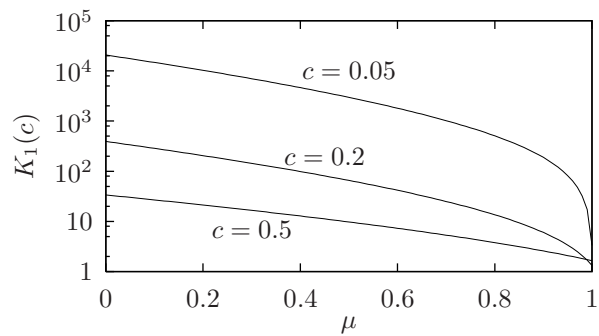}
\caption{\label{hybridgraph}The Stokes constant as a function of
$\mu$ in the model (\ref{hybridmodel}).  Note the logarithmic scale
of the vertical axis.}
\end{figure}

Finally, the last question that we would like to address here is
whether the intensity of the radiation from the moving discrete kink
depends on the type of discretization.  More specifically, we
would like to know whether the choice of one of the exceptional
discretizations (which, by definition, support translationally
invariant {\it stationary\/} kinks) serves to reduce the radiation
from the {\it moving\/} kinks.  Speight and Ward have already given
an affirmative answer for their exceptional discretization; here we
consider the one-parameter nonlinearity
\begin{equation}
Q = \frac{(1-\mu)}{2}u_n^3+\frac{\mu}{4}u_n^2(u_{n+1}+u_{n-1}),
\label{hybridmodel}
\end{equation}
which interpolates between the one-site nonlinearity
(\ref{nonlinearity-phi4}) (for which $\mu = 0$) and the exceptional
discretization (\ref{nonlinearity-Tovbis}) of Bender and Tovbis (for which
$\mu = 1$). Figure \ref{hybridgraph} shows the Stokes constant for
the model (\ref{hybridmodel}), as $\mu$ changes from 0 to 1 for
fixed values of $c$. The Stokes constant is indeed seen to be 
drastically reduced as $\mu$ approaches $1$ --- that is, in the limit of
the exceptional discretization. (It nonetheless remains
nonzero, of course, unless $c = 0$.)

\section{Concluding remarks and conclusions}

\label{section4}

In this paper we have investigated the existence of sliding kinks --- 
i.e. discrete kinks travelling at a constant velocity over a flat background, 
without emitting any radiation --- in four discrete versions of the 
quartic-coupling theory. One of these models is the most common, one-site, 
discretization. As the overwhelming majority of discrete $\phi^4$-equations,
it supports travelling kinks, but these kinks do radiate and decelerate as a
result.  The other three discretizations we considered are all 
exceptional in the sense 
that they all support one-parameter continuous
families of stationary kinks where the free parameter defines the position of the kink relative to the 
lattice. 
This property is clearly nongeneric; the translation invariance of the 
continuous $\phi^4$-theory is broken by the discretization and hence in
generic disretizations kinks may only be centred at a site or midway between 
two sites. Since the nonexistence of ``translationally-invariant'' and of sliding 
kinks in the generic models can be ascribed to similar factors, {\it viz.\/}, 
the breaking of the translation and Lorentz 
invariances, it was hoped
that the exceptional discretizations might turn out to be equally
exceptional from the point of view of 
sliding kinks. Our approach was based on the computation of the Stokes constants associated
with the putative sliding kink in a given equation.

The main conclusion of our work is that the sliding kinks do exist in the discrete $\phi^4$ theories, but
only with special, isolated, velocities
(which of course depend on $h$).
 There is one such velocity in the exceptional model of Speight and Ward,
and three different sliding velocities in the discretization  of Kevrekidis. 
It is natural to expect that the sliding kinks should play the role of 
attractors similarly to the fronts moving with ``stable velocities'' in
dissipative systems; that is, radiating travelling kinks should evolve into kinks
travelling with the sliding velocities --- if there are such velocities
in the system. Not every discretization 
supports sliding kinks, of course; in particular,  
 no sliding
velocities arise for the generic, one-site, nonlinearity 
and even for the exceptional discretization of Bender and Tovbis.

One natural way of trying to construct  the sliding kinks is via
power series expansions in powers of $c^2$; 
for the exceptional discretizations, this construction can be carried out
to any order.
This approach
was pursued in the recent work of Ablowitz and Musslimani \cite{AM03}. Our results 
indicate, however, that these power series will not converge and 
exponentially small terms 
(terms lying beyond all orders of the power expansion)
emerge because of the singular behaviour of
the Stokes
constant $K_1(c)$ as $c \to 0$. Detailed studies of this singular limit will
be presented elsewhere.

The exceptional discretizations have richer underlying symmetries than 
generic nonlinearities but the  ``translation invariance'' of the stationary kink alone
does not automatically guarantee the existence of the sliding velocities.
The exact relation between the ``translational invariance'' and mobility
of kinks is still to be clarified; at this stage it is worth mentioning that 
the Stokes constant associated with  (and hence the intensity of radiation
from) a moving kink is several orders of magnitude smaller in exceptional
models than in generic discretizations.

Finally,  it is instructive to point out some parallels with an earlier work
of Flach, Zolotaryuk and Kladko \cite{FZK99}
who also studied the phenomenon of
kink sliding in Klein-Gordon lattices. In the scheme of \cite{FZK99}, one
postulates an analytic expression for the sliding kink, $u_n(t)=\phi(n-ct-s)$,
with some explicit function $\phi(z)$,
and then reconstructs the Klein-Gordon nonlinearity for which this is an exact
solution.
Our present conclusions are in agreement with the results of these authors who
observed that for a given $h$, the kink may only slide  at a particular,
isolated, velocity.  The two approaches, ours and that of \cite{FZK99}, are
reciprocal;
while we examine the existence of sliding kinks
for  particular discretizations of the $\phi^4$-theory, with fixed parameters
independent of the kink's velocity,  in the ``inverse method'' of \cite{FZK99}
one assumes an explicit solution of a particular form but
does not have any control over the resulting nonlinearities.
Consequently, the discrete  Klein-Gordon models generated
by the ``inverse method'' are not discretizations of the $\phi^4$-theory
and do include explicit dependence on the velocity of the sliding kink.

\ack

O.O. was supported by funds provided by the South African government
and the University of Cape Town. D.P. thanks the Department of
Mathematics at UCT for hospitality during his visit and the NRF of
South Africa for financial support which made the visit possible.
I.B. is a Harry Oppenheimer Fellow; also supported by the NRF under
grant 2053723.


\section*{References}

\begin{thebibliography}{DKY05b}

\bibitem[AM03]{AM03} M.J. Ablowitz and Z.H. Musslimani,
``Discrete spatial solitons in a diffraction-managed nonlinear
waveguide array: a unified approach'', Physica D {\bf 184}, 276--303
(2003).

\bibitem[BOP05]{BOP05} I.V. Barashenkov, O.F. Oxtoby, and D.E. Pelinovsky,
``Translationally invariant discrete kinks from one-dimensional maps'',
Phys. Rev. E {\bf 72}, 035602(R) (2005).

\bibitem[BT97]{BT97} C.M. Bender and A. Tovbis, ``Continuum limit
of lattice approximation schemes'', J. Math. Phys. {\bf 38},
3700--3717 (1997).

\bibitem[DKY05a]{DKY05a} S.V. Dmitriev, P.G. Kevrekidis, and N. Yoshikawa,
``Discrete Klein-Gordon models with static kinks free of the
Peierls-Nabarro potential'', J. Phys.~A: Math.~Gen.~{\bf 38}, 7617--7627 (2005).

\bibitem[DKY05b]{DKY05b}
S.V. Dmitriev, P.G. Kevrekidis, and N. Yoshikawa,
``Standard Nearest Neighbor Discretizations of Klein-Gordon Models Cannot Preserve
Both Energy and Linear Momentum'',
 nlin.PS/0506002.

\bibitem[FZK99]{FZK99} S. Flach, Y. Zolotaryuk, and K. Kladko,
``Moving lattice kinks and pulses: an inverse method'', Phys. Rev. E
{\bf 59}, 6105--6115 (1999).

\bibitem[GJ95]{GrimshawJoshi} R.H.J. Grimshaw and N. Joshi, ``Weakly
nonlocal solitary waves in a singularly perturbed Korteweg-de Vries
equation'', SIAM J. Appl. Math. {\bf 55}, 124--135 (1995).

\bibitem[G95]{GrimshawNLS} R.H.J. Grimshaw, ``Weakly nonlocal solitary waves in
a singularly perturbed nonlinear Schr\"{o}dinger equation'', Stud.
Appl. Math. {\bf 94}, 257--270 (1995).

\bibitem[HA93]{AH93} B.M. Herbst and M.J. Ablowitz, ``Numerical chaos,
symplectic integrators, and exponentially small splitting
distances'', J. Comput. Phys. {\bf 105}, 122 (1993).

\bibitem[IJ05]{IJ04} G. Iooss and G. James, ``Localized waves in
nonlinear oscillator chains'',  Chaos {\bf 15}, 015113 (2005).

\bibitem[K03]{Panos} P.G. Kevrekidis, ``On a class of
discretizations of Hamiltonian nonlinear partial differential
equations'', Physica D {\bf 183}, 68--86 (2003).

\bibitem[KS91]{SegurKruskal2} M.D. Kruskal and H. Segur, ``Asymptotics beyond
all orders in a model of crystal growth'', Stud. Appl. Math. {\bf 85},
129--181 (1991).

\bibitem[L88]{Lazutkin} V.F. Lazutkin, ``Splitting of complex separatrices'',
Funct. Anal. Appl. {\bf 22}, 154--156 (1988).

\bibitem[PRG88]{PomeauRamani} Y. Pomeau, A. Ramani and B.
Grammaticos, ``Structural stability of the Korteweg-de Vries solitons
under a singular perturbation'', Physica D {\bf 31}, 127--134 (1988).

\bibitem[S03]{S03} J.A. Sepulchre, ``Energy barriers in coupled
oscillators: from discrete kinks to discrete breathers'', in {\em
Proceedings of the Conference on Localization and Energy Transfer
in Nonlinear Systems, June 17-21, 2002, San Lorenzo de El
Escorial, Madrid, Spain}, Eds. L. Vazquez, R.S. MacKay, M.P.
Zorzano (World Scientific, 2003), pp.~102--129.

\bibitem[SW94]{Speight1} J.M. Speight and R.S. Ward, ``Kink
dynamics in a novel discrete sine-Gordon system'', Nonlinearity
{\bf 7}, 475--484 (1994).

\bibitem[S97]{Speight2} J.M. Speight, ``A discrete $\phi^4$ system without a
Peierls-Nabarro barrier'', Nonlinearity {\bf 10}, 1615--1625
(1997).

\bibitem[S99]{Speight3} J.M. Speight, ``Topological discrete kinks'',
Nonlinearity {\bf 12}, 1373--1387 (1999).

\bibitem[TTJ98]{Tovbis1} A. Tovbis, M. Tsuchiya, and C. Jaff\'e, ``Exponential
asymptotic expansions and approximations of the unstable and
stable manifolds of singularly perturbed systems with the Henon
map as an example'', Chaos {\bf 8}, 665--681 (1998).

\bibitem[T00a]{Tovbis2} A. Tovbis, ``On approximation of stable and unstable
manifolds and the Stokes phenomenon'', Contemporary Mathematics
{\bf 255}, 199--228 (2000).

\bibitem[T00b]{Tovbis3} A. Tovbis, ``Breaking homoclinic connections for a singularly
perturbed differential equation and the Stokes phenomenon'', Stud.
Appl. Math. {\bf 104}, 353--386 (2000).

\bibitem[TP05]{TovPel} A. Tovbis and D. Pelinovsky, ``Exact conditions for
existence of homoclinic orbits in the fifth-order KdV model'',
preprint (2005).

\end{thebibliography}
\end{document}